\documentclass[fleqn,usenatbib]{mnras}

\usepackage[T1]{fontenc}

\usepackage{graphicx}	% Including figure files
\usepackage{amsmath}	% Advanced maths commands
\usepackage{amssymb}	% Extra maths symbols
\usepackage{ulem}
\usepackage{xcolor}
\definecolor{c1}{rgb}{178,24,43}
\definecolor{c2}{rgb}{239,138,98}
\definecolor{c3}{rgb}{253,219,199}
\definecolor{c4}{rgb}{209,229,240}
\definecolor{c5}{rgb}{103,169,207}
\definecolor{c6}{rgb}{33,102,172}

\newcommand{\gram}{\mbox{\,g}}
\newcommand{\cs}{c_{\rm s}}

\newcommand{\kms}{\mbox{\,km s}^{-1}}

\newcommand\Msun{M_\odot}
\newcommand{\muG}{\,\mu\mbox{G}}
\newcommand{\Myr}{\mbox{\,Myr}}
\newcommand{\pc}{\mbox{\,pc}}
\newcommand{\pcc}{\mbox{\,cm}^{-3}}
\newcommand{\talf}{\tau_{\rm A}}
\newcommand{\td}{\tau_{\rm d}}
\newcommand{\tff}{\tau_{\rm ff}}
\newcommand{\valf}{v_{\rm A}}

\title[A time-scale based resolution criterion]{A resolution criterion based on characteristic time-scales for MHD simulations of molecular clouds}

\author[Granda-Mu\~noz et al.]{Guido Granda-Mu\~noz,$^{1}$\thanks{E-mail: g.granda@irya.unam.mx}
Enrique V\'azquez-Semadeni,$^{1}$\thanks{E-mail: e.vazquez@irya.unam.mx}
Gilberto C. G\'omez,$^{1}$\thanks{E-mail: g.gomez@irya.unam.mx} and  
\newauthor
Manuel Zamora-Avil\'es$^{2}$
\\
$^{1}$Instituto de Radioastronom\'ia y Astrof\'isica, Universidad Nacional Aut\'onoma de M\'exico, Apdo. Postal 3-72, Morelia, Michoac\'an 58089, M\'exico\\
$^{2}$CONACYT-Instituto Nacional de Astrof\'isica, Optica y Electr\'onica, Luis E. Erro 1, 72840 Tonantzintla, P\'uebla, M\'exico
}

\date{Accepted 2021 December 9. Received 2021 December 7; in original form 2021 October 8}

\pubyear{2021}

\begin{document}
\label{firstpage}
\pagerange{\pageref{firstpage}--\pageref{lastpage}}
\maketitle

\begin{abstract}
We investigate the effect of numerical magnetic diffusion in magnetohydrodynamic (MHD) simulations of magnetically supported molecular clouds. 
To this end, we have performed numerical studies on adaptive mesh isothermal simulations of marginally subcritical molecular clouds.
We find that simulations with low and intermediate resolutions collapse, contrary to what is theoretically expected. However, the simulation with
the highest numerical resolution oscillates around an equilibrium state without collapsing. 
In order to quantify the numerical diffusion of the magnetic field, we ran a second suit of current-sheet simulations in which the numerical magnetic diffusion coefficient can be directly measured and computed the corresponding diffusion times at various numerical resolutions. On this basis, we propose a  criterion for the resolution of magnetic fields in MHD simulations based on requiring that the diffusion 
time to be larger than the characteristic time-scale of the physical process responsible for the dynamic evolution 
of the structure. 
\end{abstract}

\begin{keywords}
magnetic fields -- MHD -- methods: numerical -- magnetic fields -- ISM: clouds -- ISM:magnetic fields.
\end{keywords}

%%%%%%%%%%%%%%%%%%%%%%%%%%%%%%%%%%%%%%%%%%%%%%%%%%

%%%%%%%%%%%%%%%%% BODY OF PAPER %%%%%%%%%%%%%%%%%%
\noindent
\section{Introduction}

Magnetohydrodynamic simulations of the formation and evolution 
of molecular clouds have played an important role in the study of molecular clouds and star formation,
either for the magnetic support model,
where magnetic fields are responsible for supporting the clouds and their substructures against gravitational collapse
\citep{shuStarFormationMolecular1987,mouschoviasCosmicMagnetismBasic1991},
or for the turbulent support model, where the main support mechanism is the dynamical pressure generated by turbulence \citep{elmegreenStarFormationCrossing2000,maclowControlStarFormation2004}.
Therefore, it is important to study the effects of numerical dissipation of
the magnetic field on magnetohydrodynamic simulations of 
molecular clouds and their substructure. 

One common way of measuring the importance of magnetic fields in the support of molecular clouds and their
substructure is by computing the mass-to-flux ratio, denoted by $\mu$  when normalized to a critical value for marginal support, which depends on the geometry, but is otherwise a constant. When a cloud
is magnetically supported against gravitational collapse, it is
referred to as subcritical and, theoretically, $\mu <1 $. In this case, the cloud is expected to contract partially and then attain a hydrostatic configuration, flattened along the field direction \citep{mouschoviasNoteCollapseMagnetic1976}.
Otherwise, the cloud is said to be supercritical, undergoes gravitational collapse, and has $\mu >1$ \citep{mestelStarFormationMagnetic1956}. However, if the magnetic field is insufficiently resolved, it is possible that a simulation which is in principle subcritical may nevertheless undergo spurious gravitational collapse due to numerical diffusion of the magnetic flux.

Resolution criteria are necessary for the adequate numerical simulation of every physical process. For example, \cite{trueloveJeansConditionNew1997} found that resolving
the Jeans length with a minimum of four cells avoids spurious fragmentation
on AMR hydrodynamic simulations of isothermal molecular clouds. Additionally, \cite{bateResolutionRequirementsSmoothed1997} found a similar resolution condition for SPH simulations. Also, \cite{koyamaFieldConditionNew2004}  found that at least three cells 
are necessary to resolve the Field length and to achieve convergence of some properties 
such as the number of clouds formed by thermal instability and the maximum Mach number in simulations of the development of turbulent motions driven by the
non-linear evolution of thermal instability. 
More recently, \cite{federrathNewJeansResolution2011} studied the gravity-turbulence-driven magnetic field amplification of supercritical clouds. They found that
it is necessary to resolve the Jeans length with at least 30 cells in order to resolve turbulence at the Jeans scale and capture minimum dynamo  amplification of the magnetic field.

Resolution criteria are generally obtained by means of convergence tests, which consist of increasing the resolution until a certain feature of the flow remains invariant as the resolution is increased. In this sense, convergence tests constitute a trial-and-error procedure.
In this work, we propose instead a resolution criterion based on measuring a ``numerical diffusion coefficient'' via a test problem, from which the dependence of the numerical diffusion time-scale on the resolution can be inferred and compared with the characteristic timescale of the physical process being investigated, thus providing a physically-motivated prescription for the necessary resolution.
Specifically, the requirement is
 that the diffusion time needs  to be longer than the relevant dynamical time of the structure. In this paper, we present an application to the problem of adequately resolving the magnetic support against the self-gravity of a dense molecular cloud core.

The paper is organized as follows. In section \ref{sec_mc}, we present a  suite of numerical simulations of marginally magnetically subcritical  molecular clouds at various resolutions, which undergo spurious collapse when the magnetic field is insufficiently resolved. In section \ref{resolution_sect}, we propose a resolution
criterion based on estimating the physical characteristic time-scale of the physical process being simulated, a measurement of the numerical diffusion coefficient by means of a test simulation, and apply it to the problem of magnetic cloud support. In section \ref{harris_sect}, the suite of simulations aimed at computing 
the numerical magnetic diffusion coefficient is presented. In section \ref{results_sect}, we derive the numerical magnetic diffusion coefficient and apply our criterion to the magnetic cloud support simulations, finding an agreement with the resolution needed to resolve the support. In section \ref{discussion_sect}, we discuss the implications of our results
and present our conclusions.

\section{Molecular cloud simulations}\label{sec_mc}

\subsection{ Numerical set-up} \label{sec_num_setup}
In this section, we present a suite of numerical simulations of magnetic support in molecular clouds  using the adaptive mesh refinement (AMR) code {\sc{flash}}, version 4.5, \citep{fryxellFLASHAdaptiveMesh2000,2008PhST..132a4046D,DUBEY2009512}, and the ideal magnetohydrodynamic (MHD) multi-wave HLL-type solver \citep{waaganRobustNumericalScheme2011}. The gravitational solver applied 
for these simulations is the OctTree algorithm also included in {\sc{flash}} \citep{2018MNRAS.475.3393W}, while for the adaptive refinement, we use 
the L\"{o}hner's error estimation applied to the density \citep[]{1987CMAME..61..323L}.

We consider an isothermal cloud that
is marginally supported by the magnetic field
(i.e. marginally subcritical), and find the minimum numerical resolution necessary for the cloud to actually be supported, rather than collapsing due to the loss of magnetic support caused by numerical diffusion. Each of these simulations has the same initial conditions and starts with an initial effective resolution of 32 cells per dimension, although each simulation reaches a different maximum resolution (see Table \ref{tab:mc2}). They also include periodic boundary conditions for both the magnetohydrodynamics and the self-gravity.

\begin{table*}
 \caption{Molecular cloud simulations. In the first, second and third columns the simulation's name, it's effective and maximum resolution is shown. In the fourth column, we show the magnetic pressure gradient magnitudes computed at half the Jean's length. The diffusion 
 coefficients, free-fall times , diffusion times and collapse times at the dynamically equivalent time (see text) are shown in the fifth, sixth, seventh and eight columns respectively.}
 \label{tab:mc2}
 \begin{tabular}{cccccccc}
 \hline
 Simulation&Effective resolution&Maximun resolution $(\pc)$ &${|\nabla P_{B}|} \ \mathrm{(dyn/cm^{3})}$ & $\eta \ (\pc^{2}/\Myr)$ & $\tau_{ff}(\Myr)$ & $\tau_{d}\ (\Myr)$ & $\tau_{coll} \ (\Myr) $\\\hline
 MC7 & 128 &$\mathrm{7.81\times 10^{-2}}$ & $\mathrm{1.565\times 10^{-29}}$ & 0.293 & 0.428 & 0.083 & 9.6 \\
 MC8 & 256 &$\mathrm{3.91\times 10^{-2}}$  &$\mathrm{1.041\times 10^{-29}}$& 0.036 & 0.353 & 0.384 & 9.8 \\
 MC9 & 512 &$\mathrm{1.95\times 10^{-2}}$ &$\mathrm{1.218\times 10^{-30}}$&  0.001 & 0.697 & 51.907 & -- \\\hline
 \end{tabular}
\end{table*}

The initial conditions consist of an initially uniform magnetic field 
along the $x$-axis of $ 25.17 \muG$, a box size of $10 \pc$, and a density 
perturbation of a 3D Gaussian profile on top of a background density $\rho_{0}$:

\begin{equation}
 \centering
 \label{eqfirst}
   \rho = \rho_{0} \Bigr \{1+A\ \exp\Bigr[-\frac{1}{2\sigma^{2}}\Bigr( x^{2} + y^{2} +z^{2} \Bigr)\Bigr]\Bigr \}, 
\end{equation}

\noindent
where $A$ and $\sigma$ are constants which represent the perturbation amplitude and a measure of the perturbation size respectively. For this simulation,
$\rho_{0}=\mathrm{2\ldotp12\times 10^{-22}\gram\pcc}$, $A =1.50$, $\sigma = 2.5$ pc. This set-up results in a total mass in the computational volume of $4.14\times 10^{3} \Msun$,
a Jeans' length of $2.66 \pc$, given by
\begin{equation}
 \centering
 \label{eq_jeans}
 \lambda_{J}=\sqrt{\frac{\rm{\pi} \cs^{2}}{G\rho}},
\end{equation}
where $c_{s}=0.2\kms$ is the sound speed, and a mass-to-flux ratio $\mu = 0.53$
where we considered spherical geometry.

It is important to notice that this estimated value for the 
mass-to-flux ratio ($\mu=0.53$) yields in practice a marginally subcritical molecular cloud because
it does not include the contribution of the external gas and magnetic pressures on the computation of the critical mass-to-flux ratio \citep{shuPhysicsAstrophysicsVolume1992} and/or the flat geometry of the cloud due the mass accretion along the magnetic field lines \citep{strittmatterGravitationalCollapsePresence1966}. The  marginal nature of the subcritical condition is an important feature, since otherwise the cloud evolution should not be very different for simulations with similar numerical resolutions.

\begin{figure*}
\includegraphics[width=\textwidth]{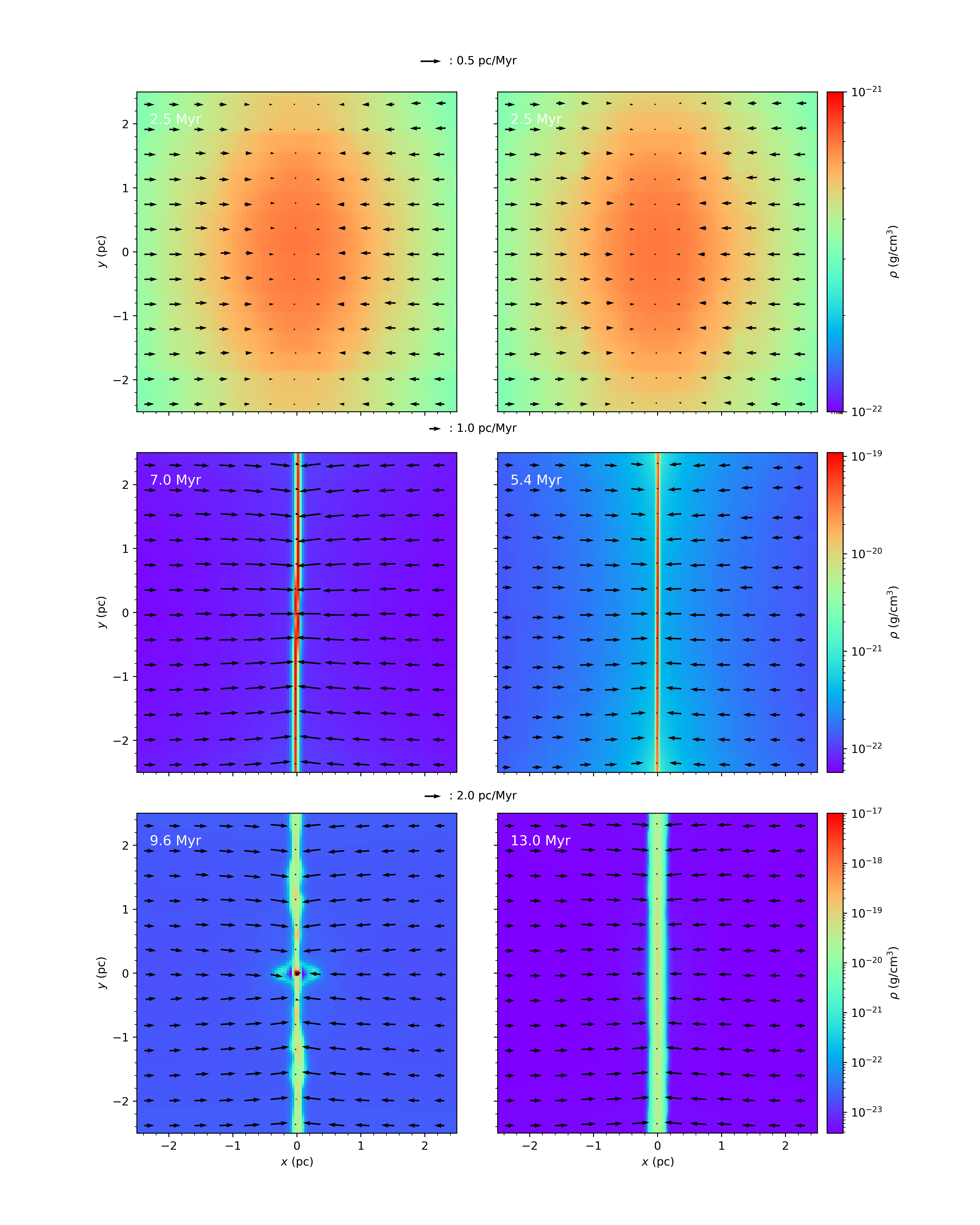}
    \caption{Zoomed cross section through the $z=0$ plane. Colours represent gas density and the velocity field is represented by the black arrows. Left column correspond to the MC8 simulation while right column to the MC9 simulation. Note the different evolution times for each simulation. Simulation MC8 has already developed a collapsed object at its center by $t=9.6$ Myr, while run MC9 has not done it even by $t=13.0$ Myr.}
    \label{fig:1}
\end{figure*}

\subsection{Results of the simulations} \label{sec_num_res}

As mentioned in the previous section, the three simulations have identical physical parameters and differ only in the maximum resolution allowed in each of them. We find that, while the low (MC7) and medium resolution (MC8) simulations undergo 
collapse after $9.6$ and $9.8 \Myr$, respectively, a dense structure
is formed and oscillates around an equilibrium state for the high-resolution run (MC9). 

\begin{figure}
\includegraphics[width=\columnwidth]{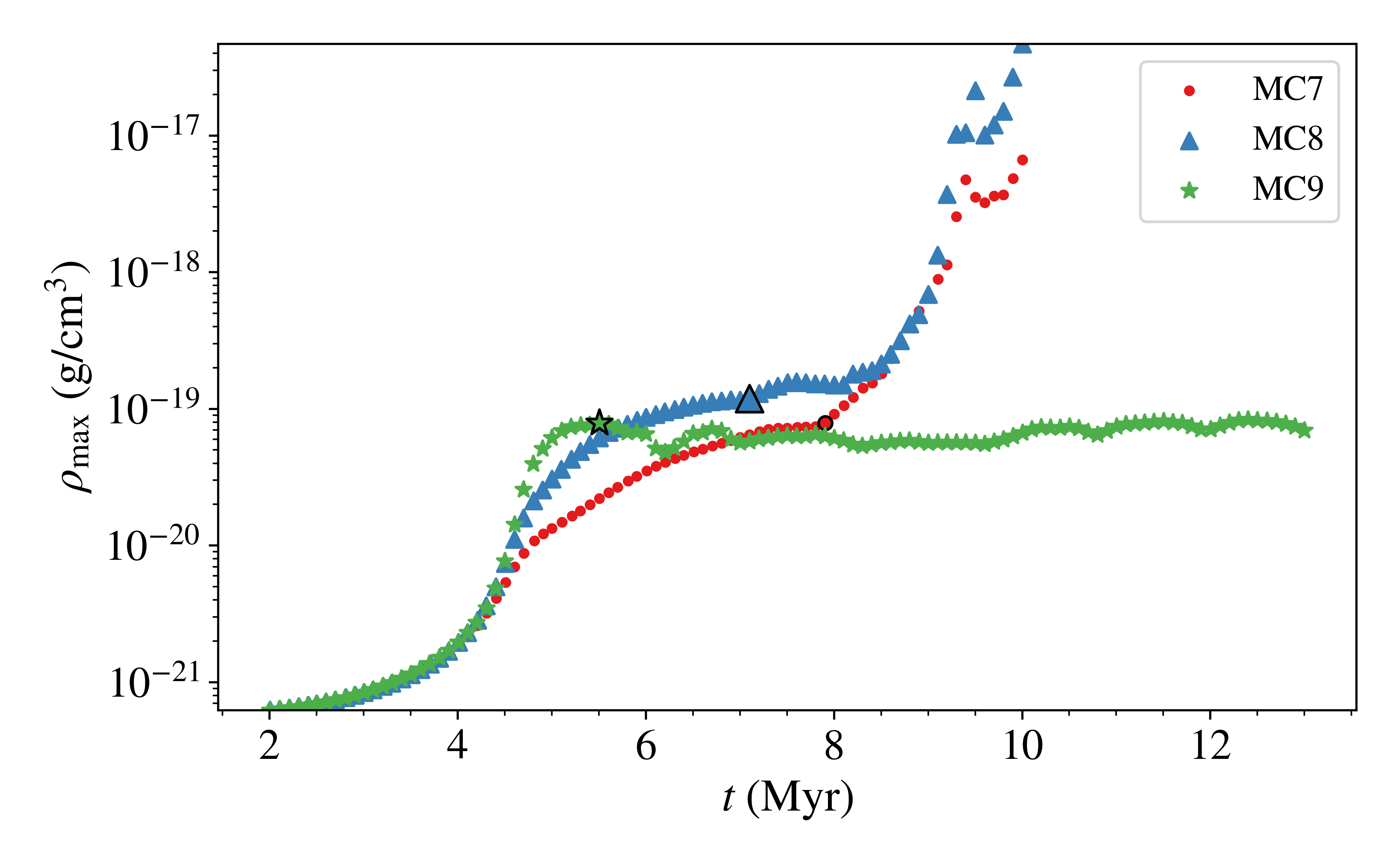}
    \caption{Temporal evolution of the maximum density in each of the MC simulations. The dark edge-colored $\textcolor[RGB]{228,26,28}{\bullet}$
    $\textcolor[RGB]{55,126,184}{\blacktriangle}$, $\textcolor[RGB]{77,175,74}{\bigstar}$
    represent the dynamical equivalent densities for the MC7, MC8, and MC9 simulations respectively.}
    \label{fig:2}
\end{figure}

The evolution of the intermediate (MC8) and high-resolution (MC9) simulations are shown in Fig. \ref{fig:1}.\footnote{This and the other plots present in this manuscript were done using yt project \citep{turk2011}.} Both simulations start
their evolution in a very similar way (top left and top right panels), collecting
gas on the central part of the computational domain along the magnetic 
field direction. The maximum density in both simulations increases and eventually reaches a plateau on its temporal evolution
which is shown in Fig. \ref{fig:2}. The time to reach the plateau density value is different in each simulation. We attribute this to the presence of different levels of numerical diffusion at each resolution, so that magnetic support is lost more rapidly at lower resolution. Therefore, we consider a different, but dynamically equivalent, time  in each simulation, defined as the time at which the maximum density reaches the plateau stage on its temporal evolution. We call this stage the dynamically equivalent for the different simulations. 
The dynamically equivalent times are $7.8$, $7.0$, and $5.4 \Myr$  for the simulations MC7, MC8, and MC9 respectively, and we consider them as the computation snapshots for the diffusion and dynamical times estimates (see Section \ref{sect_dy_times}).

After the dynamically equivalent stage, the final evolution state of each simulation depends on whether the resolution was enough for solving the magnetic field correctly or not. For MC8, as shown in the bottom left panel of Fig. \ref{fig:1}, the structure collapses due to 
the poor numerical resolution, or equivalently, to its high numerical magnetic diffusion coefficient. On the contrary, as shown in the bottom right panel of this figure, MC9 oscillates around an equilibrium state
without collapsing in agreement with its subcritical condition. Therefore, numerical diffusion of the magnetic field can cause a spurious collapse of a marginally sub-critical molecular cloud. 

\section{Resolution criterion} \label{resolution_sect}

In the previous section, we empirically found the resolution necessary to
properly resolve the magnetic support of a subcritical molecular cloud. However, it would be desirable to have a more physically motivated, predictive criterion which can then be used in simulations in regard to effects other than the magnetic field.
Since the evolution of any physical process is controlled
by its relevant dynamical time, we propose a resolution criterion based on comparing the numerical diffusion
time in a given simulation to the dynamical time of the physical process under investigation.

For subcritical molecular clouds, we obtain the relevant dynamical
times from the condition $\mu<1$, which requires that the Alfv\'en 
crossing time be less than the free-fall time:
\begin{equation}
 \label{eq0}
  \talf < \tff,
\end{equation}
where the Alfv\'en crossing time is given by
\begin{equation}
    \label{eq1}
     \talf = L/\valf,
\end{equation}
with $L$ being the relevant spatial scale and $\valf$ the Alfv\'en velocity.
In turn, the free-fall time is given by
\begin{equation}
 \label{eq3}
 \tff =  \Bigr (\frac{ 3 \pi}{ 32 G \rho }\Bigr)^{1/2} ,
\end{equation}
%\noindent
where $G$ is the gravitational constant. To obtain the resolution criterion, we compare the diffusion time with these time scales. 

The diffusion time in the simulation may be obtained using the fact that the effect of the spatial discretization due to the numerical grid on the evolution of magnetic fields can be computed in terms of a magnetic diffusivity coefficient $\eta$ \citep[e.g.][]{bodenheimerNumericalMethodsAstrophysics2007}.
Hence, a diffusion time may be computed in terms of $\eta$,
\begin{equation}
 \label{eq4}
 \td =L^{2}/\eta,
\end{equation}
%
%\noindent
where $\eta$ is the numerical magnetic diffusion coefficient. So,
a smaller $\eta$ means a larger diffusion time and ideal MHD is achieved when it is equal to zero.

When the numerical magnetic diffusion is large and controls the dynamics
of the structure, we have
\begin{equation}
 \label{eq5}
  \td < \talf.
\end{equation}
In this case, the evolution of the structure is not physical because it is driven by numerical diffusivity
and the Alfv\'en waves up to the wavelength for which $\td = \talf$ are damped by numerical diffusion.

Increasing the numerical resolution reduces the numerical magnetic diffusion coefficient, so, the numerical magnetic diffusion time given by equation \eqref{eq4} increases. Therefore, the numerical
resolution should be increased until the numerical diffusion and dynamical times fulfill the relation
\begin{equation}
 \label{eq7}
 \tff , \talf <\td. 
\end{equation}
When relation \eqref{eq7} is satisfied, the diffusion time is larger 
than the dynamical time of the mechanism responsible for the 
support of the molecular cloud, namely the propagation of MHD waves, for which the relevant time-scale is the Alfv\'en crossing time. Therefore, when relation (\ref{eq7}) is satisfied, Alfv\'en waves can propagate without significant numerical diffusion during a free-fall time, \footnote{Note that we will actually
consider {\it twice} the free-fall time as the relevant time-scale because numerical simulations consistently show this to be the order of the actual collapse time, since the thermal pressure gradient is not negligible during the first stages of the collapse \citep[e.g.,] [] {Larson69, Galvan+07, naranjo-romero_hierarchical_2015}.} 
 and thus our proposed resolution criterion based on the characteristic time-scales of the problem consists in finding a numerical resolution which ensures that relation \eqref{eq7} is
satisfied. The problem becomes now the estimation of the numerical diffusion time-scale as a function of resolution.
 
\section{Harris-like current-sheet simulations}\label{harris_sect}

In order to apply the resolution criterion given by relation \eqref{eq7}, we first need to
measure the numerical magnetic diffusion coefficient $\eta$. With this in mind, 
we simulate a Harris-like 
current-sheet \citep[e.g.][]{skala3DMHDCode2015,kliemSolarFlareRadio2000}. This simulation consists in setting up a magnetic field configuration that reverses direction across a narrow region, maintaining total (thermal + magnetic) pressure equilibrium. Following the set-up of \cite{skala3DMHDCode2015}, these
simulations are two dimensional and isothermal with a computational 
domain of $[-5,\ 5] \pc$ on the $x$-axis, $[-0.6,\ 0.6] \pc$ on the $y$-axis, and open and 
periodic boundary conditions in the $x$ and $y$ directions, respectively. Note that these simulations do not include self-gravity.

The initial density and magnetic field intensity are given by
\begin{equation}
 \label{eq9}
 \rho=\left(P_{\rm tot} - P_{B,\rm par}\ \tanh^2(x)\right)/\cs^{2},
\end{equation}
\begin{equation}
 \label{eq10}
  B_{y}=\left(8\pi \ P_{B,\rm par}\right)^{1/2} \tanh(x),
\end{equation}
where $P_{\rm tot} = P_{\rm th} + P_B$ is the total pressure, $P_{\rm th} = \cs^2 \rho$ is the thermal pressure, and $P_B = B^2/8 \pi$ is the magnetic pressure. $P_{B,\rm par}$ is the asymptotic value of the magnetic pressure at large $x$. 

Numerical magnetic diffusion disrupts the initial pressure equilibrium 
in the central region of the computational domain, which is the region
where the gradient of the magnetic field is the largest. In order to measure
the magnetic diffusion coefficient on this region, we consider the induction equation in the presence of resistivity: 
\begin{equation}
  \label{eq11b}
  \frac{\partial\bmath{B}}{\partial t} + \nabla \times (\bmath{B}\times \bmath{v}) = - \nabla \times (\eta \nabla \times \bmath{B}),
\end{equation}
where $\bmath{v}$ is the fluid velocity, which, for our pressure equilibrium configuration, is zero. Assuming that $\eta$ is uniform in space, we obtain
\begin{equation}
 \label{eq12}
 \eta =  \frac{\Bigr (\frac{\partial B} {\partial t} \Bigr )} {\Bigr (\frac{\partial^{2} B} {\partial x^{2}} \Bigr)}. 
\end{equation}
Hence, in order to measure the numerical resistivity corresponding to a given resolution, we performed the suite of simulations described in Table \ref{tab:harris2}, in which the derivatives appearing in eq.\ (\ref{eq12}) are to be measured. Each 
of the simulations has the same initial conditions but different numerical resolution. The simulations were performed with the same version of the {\sc{flash}} code and MHD solver.

\begin{figure}
\includegraphics[width=\columnwidth]{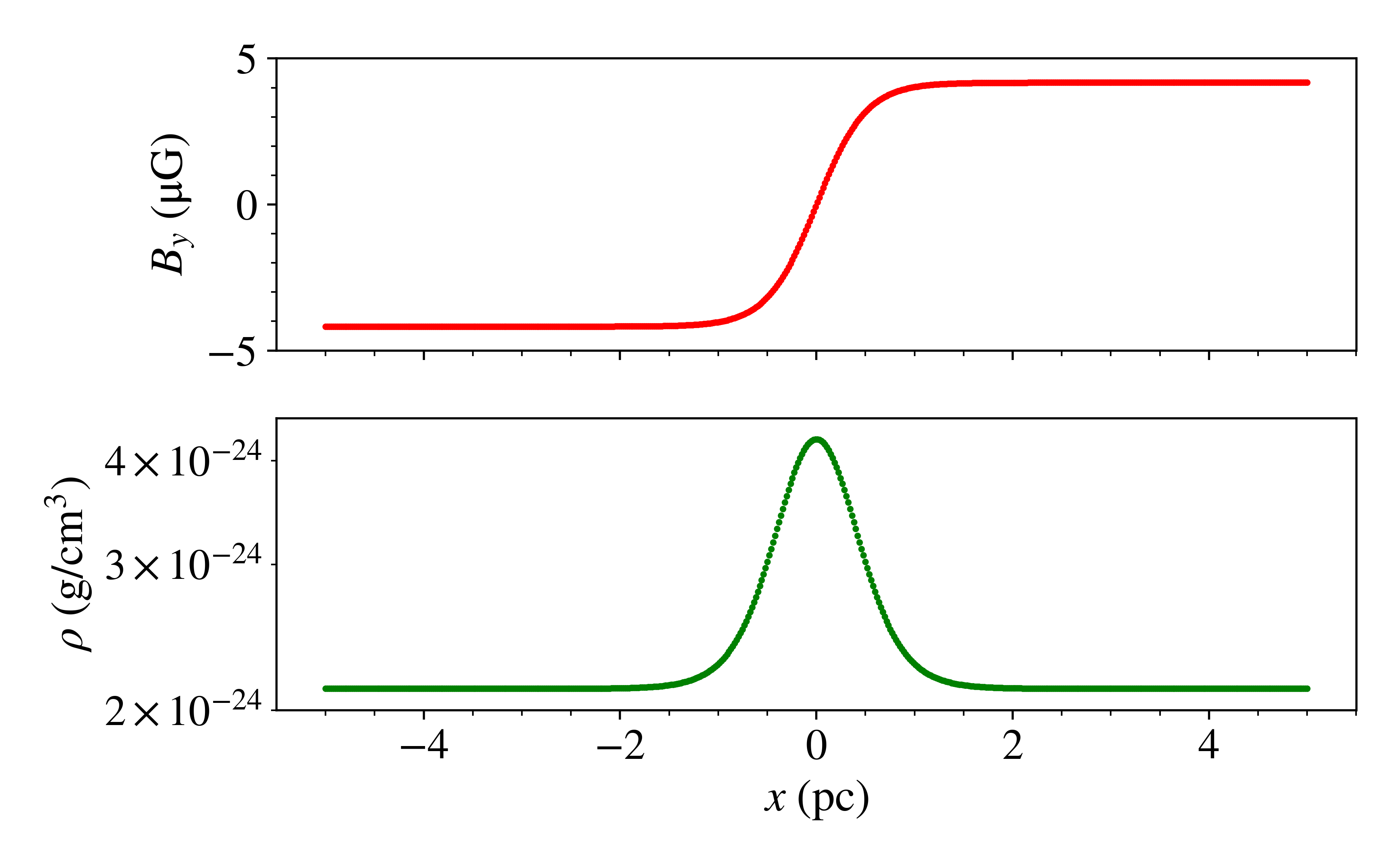}
    \caption{Magnetic field and density initial conditions for the Harris-like simulations.}
    \label{fig:ic}
\end{figure}

\section{Results} \label{results_sect}
\subsection{Numerical magnetic diffusion coefficients of the Harris-like simulations}\label{eta}

\begin{figure}
\includegraphics[width=\columnwidth]{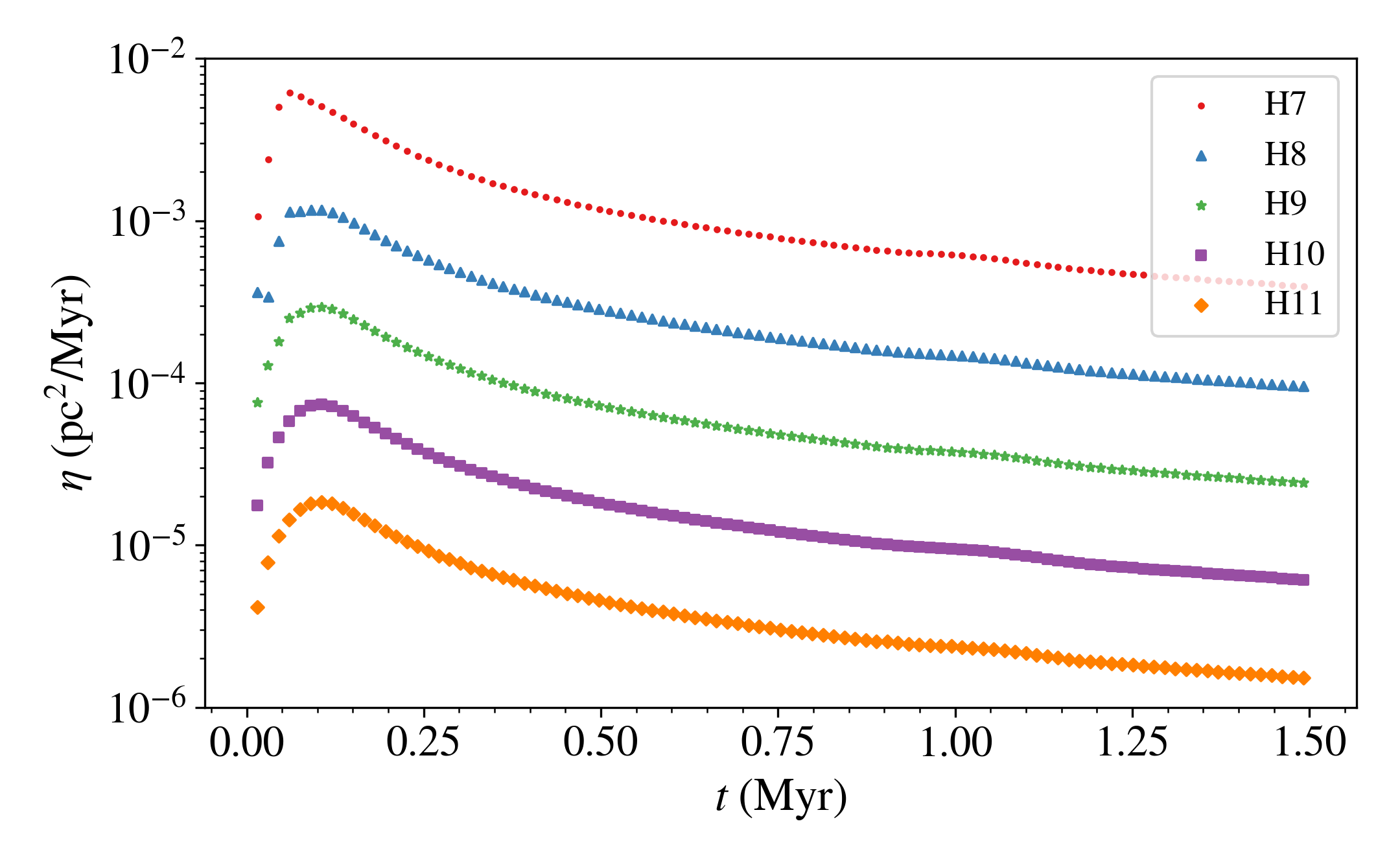}
    \caption{$\eta$ {\it versus} time for the Harris-like simulations.}
    \label{fig:3}
\end{figure}

To measure the numerical magnetic diffusion coefficients of each of the 
simulations presented in Table \ref{tab:harris2}, where we have set the initial conditions to obtain the density and magnetic field profiles 
shown in Fig. \ref{fig:ic}. We evaluated numerically 
equation \eqref{eq12} on the point $({x_{\rm m}, \ y_{\rm m}) = (-0.5493\, {\rm pc},0)} \equiv (- x_0,\ 0)$,
which is the point where the magnetic field strength is half of its maximum magnitude.
The magnetic diffusion coefficient measured over time, following the prescription by \citet{skala3DMHDCode2015}, is shown 
in Fig. \ref{fig:3}. As expected, a higher resolution 
yields a smaller numerical magnetic diffusion coefficient, thus a larger diffusion time.

Since $\eta$ varies in time, we consider its maximum  value at each resolution as the measured value to avoid 
that the time term present on the denominator 
of the discretization of equation \eqref{eq12} dominates its temporal evolution. The resulting numerical diffusion coefficients are listed in the third column of Table \ref{tab:harris2}.\footnote{We also computed 
the diffusion parameters for the eight-wave MHD solver included in the {\sc{flash}} 4.5 distribution. 
These results are shown in Appendix \ref{appendix}}
%obtaining less diffusive values ranging 
%from 68.31\% to 98.39\% of the ones shown in Table \ref{tab:harris2} for the H7 and H9 simulations, respectively.}

\begin{table}
 \caption{Harris-like simulations. The first and second columns represent the simulation name and its resolution respectively. The third and fourth columns
 represent numerical magnetic diffusion coefficients and magnetic pressure gradient magnitudes.}
 \label{tab:harris2}
 \begin{tabular}{cccc}
 \hline
 Simulation & Resolution (\pc) &$\eta \ (\pc^{2}/\Myr)$  & $ |\nabla P_{B}| \ (\mathrm{dyn/cm^{3}})$ \\\hline
 H7 & $\mathrm{7.81\times 10^{-2}}$ &$\mathrm{6.164\times 10^{-3}}$& $\mathrm{3.289 \times 10^{-31}}$\\
 H8 & $\mathrm{3.91\times 10^{-2}}$ &$\mathrm{1.170\times 10^{-3}}$& $\mathrm{3.406 \times 10^{-31}}$\\
 H9 &  $\mathrm{1.95\times 10^{-2}}$ &$\mathrm{2.950\times 10^{-4}}$& $\mathrm{3.393\times 10^{-31}}$\\
 H10 &  $\mathrm{9.75\times 10^{-3}}$ &$\mathrm{7.399\times 10^{-5}}$& $\mathrm{3.383\times 10 ^{-31}}$\\
 H11 & $\mathrm{4.88\times 10^{-3}}$ &$\mathrm{1.849\times 10^{-5}}$& $\mathrm{3.377\times 10^{-31}}$\\\hline
 \end{tabular}
\end{table}

\subsection{Scaling of \pmb{$\eta$} for different conditions}\label{eta_scale}

In addition to depending on the resolution, the value of the numerical magnetic diffusion coefficient depends also on 
the magnetic field derivatives in the region where it is measured. %, as can be seen from equation \eqref{eq12}.
In order to estimate this dependence,
we performed several Harris-like current
sheet simulations for a range of values of the magnetic pressure parameter $P_{B,\rm par}$ (see eq.\ \ref{eq10}) while keeping
the same numerical resolution and the same width of the central transition region (i.e., keeping the same value of $x_0$) used in simulation H7. Thus, varying $P_{B,\rm par}$ is equivalent to varying the magnetic pressure gradient, since $x_0$ is kept constant.

\begin{figure}
\includegraphics[width=\columnwidth]{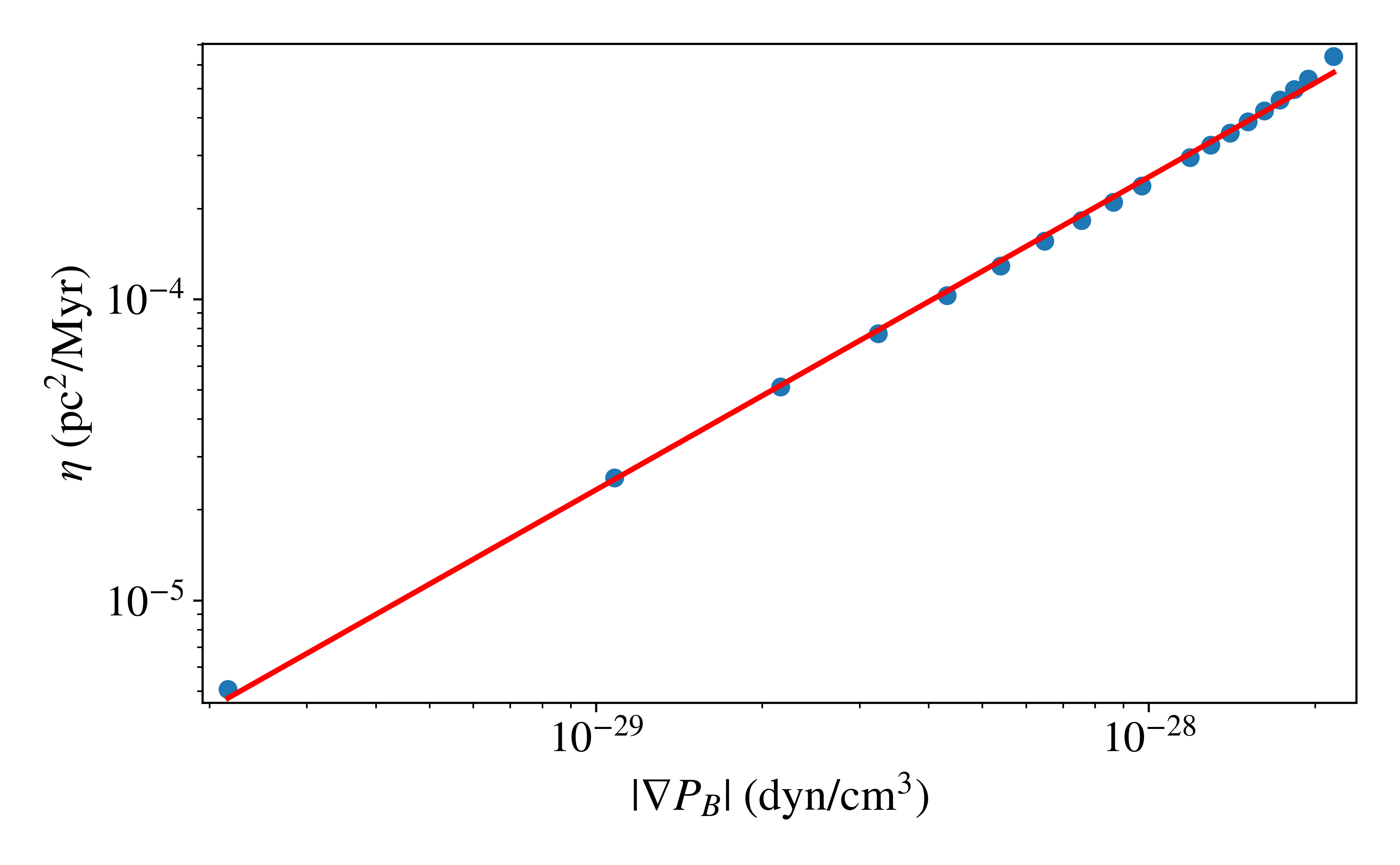}
    \caption{The numerical diffusivity $\eta$ {\it versus} the magnetic pressure gradient magnitude. The blue points represent the measured values
    while the red line is a fit to those points.}
    \label{fig:4}
\end{figure}

For this set of simulations, we chose values of $P_{B,\rm par}$ that correspond to magnetic field values in the range of [10.56, 105.64] $\mu G$.  
We find that $\eta$ scales almost linearly with the magnetic pressure gradient, as shown in Fig. \ref{fig:4}, for which we obtained a fit given by $\log(\eta) \approx 25.48 + 1.04 \log(|\nabla P_{B}|)$. We attribute this behavior to the fact that, in the Harris simulations, the driver of the numerical diffusion is the magnetic pressure gradient. Therefore, this result allows us to incorporate different values of the magnetic field gradient into the computed value for $\eta$ by making the correction

\begin{equation}
 \label{eq12.1}
 \eta_{2} = \eta_{1} \frac{|\nabla P_{B2}|}{|\nabla P_{B1}|},
\end{equation}
where $\eta_{2}$ and $\eta_{1}$ are the numerical magnetic diffusion coefficient under the conditions
of magnetic pressure gradient $\nabla P_{B2}$ and $\nabla P_{B1}$ respectively.
 In our case, this allows us to use the derived values for $\eta$ in the Harris-like simulations. We, therefore, proceed as follows: For a given refinement level, the pressure gradient, $\nabla P_{B1}$, is measured at the same point as $\eta_{1}$ (see section \ref{eta}) in the Harris-like simulations. These measurements are listed in the third column of Table \ref{tab:harris2}. In the molecular cloud simulations, $\nabla P_{B2}$ is measured
at a distance equal to half the Jeans length from the cloud's center, since we are studying the evolution of a molecular
cloud that would collapse in the absence of magnetic support, therefore, the Jeans length is the relevant characteristic length scale. Therefore, using the measured values for $\eta_{1}$, |$\nabla P_{B1}$|, |$\nabla P_{B2}$| (listed in Tables \ref{tab:harris2} and \ref{tab:mc2}), as well as equation \eqref{eq12.1}, 
we obtain the values of $\eta_{2}$ also listed in Table \ref{tab:mc2}). In turn, this allows
us to compute the diffusion time and compare it with the free-fall time to ensure 
that the numerical resolution is enough to correctly model the dynamical evolution.

\subsection{Diffusion and dynamical times}\label{sect_dy_times}
According to eq.\ \eqref{eq4}, the diffusion time-scales can be obtained from the numerical magnetic diffusion coefficients and
the spatial scale across which the magnetic field is being diffused. As mentioned in the previous subsection, we consider half the Jeans length as the diffusion length scale. The diffusion and free-fall times are computed at the time when the simulations reach their dynamically equivalent state (see Section \ref{sec_mc}) are listed in Table \ref{tab:mc2}. It is worth noting that the computed free-fall times
differ only slightly for the different resolution simulations, while the estimated diffusion times vary by almost 3 orders of magnitude, as a consequence of the strong dependence of the magnetic diffusion coefficient on numerical resolution.
Specifically, for the MC7 and MC8 simulations,
the diffusion time is smaller than twice the free-fall time. In other words, 
the numerical diffusion of the magnetic field controls the dynamics of these clouds.
Unsurprisingly, the subcritical cloud spuriously collapses.

In contrast, for the MC9 simulation, the diffusion time is larger than twice the free-fall time. So, this simulation is not dominated by the numerical diffusion of the magnetic field and it does not collapse. 

In conclusion, the results from this section show that our resolution criterion, based on comparing the numerical diffusion time scale with the characteristic time-scale of the physical problem, is consistent with the resolution empirically found to be necessary in order to correctly simulate 
the magnetic support of a cloud in Section \ref{sec_mc}.

\section{Discussion and conclusions} \label{discussion_sect}

In this work, we have found that a numerical simulation of a marginally magnetically subcritical molecular cloud undergoes spurious collapse if the numerical resolution is insufficient, and presented two different approaches to estimate the resolution required in order to properly resolve the magnetic support. The first one consisted in a study of the evolution of a marginally sub-critical molecular
cloud, 
finding that when the resolution is poor, numerical
diffusion of the magnetic field causes the spurious collapse of the cloud. 
The second approach consisted in the implementation of a physically motivated resolution criterion, relying on the comparison of the numerical magnetic diffusion time implied by the resolution used with the relevant
dynamical time, in this case, (twice) the free-fall time.

This criterion recovers the required resolution, but in addition, it provides a physical interpretation and a {\it predictive} procedure for the choice of the required resolution, provided some additional numerical tests are performed in order to estimate the numerical diffusion for a given physical  process, using a given solver.
For example, for cold atomic clumps that can grow from thermal instabilities  in the presence of a magnetic field on time-scales of the order of the cooling time, we may compare this dynamical timescale with the magnetic diffusion time
and obtain equations analogous to \eqref{eq5} and \eqref{eq7}.

It is important to note that, in this work, we have restricted our study to the effect of numerical magnetic diffusion on the evolution and collapse of subcritical molecular clouds.
On the other hand, in the case of supercritical clouds with $\mu>1$, the relation between the free-fall and the Alfv\'enic crossing times
corresponding to that presented in equation \eqref{eq0} becomes instead
\begin{equation}
\label{eq13}
\tff< \talf.
\end{equation}
In this case, collapse always occurs but, if the resolution is insufficient, the collapse may occur too rapidly, since it is known that the magnetic forces can in principle delay it \citep[e.g.,] [] {Ostriker+99}. Thus, the required
numerical resolution to avoid this situation is the one that ensures the fulfillment of
\begin{equation}
 \label{eq14}
 \tff <\talf<\td.
\end{equation}
Therefore, insufficient resolution in either the subcritical or supercritical cases may lead to an overestimation of the star formation rate.

\section*{Acknowledgements}
This research was supported by a CONACYT scholarship. We are thankful
to Adriana Gazol and Jos\'e Juan Gonz\'alez Avil\'es for
their helpful comments and feedback, to Robi Banerjee for providing the MHD solver, and to the referee, James Beattie, for useful comments.
\section*{Data availability}
The data underlying this article will be shared on reasonable request to the corresponding author.
\bibliographystyle{mnras}
\bibliography{paper_collapse} % if your bibtex file is called example.bib

%\end{thebibliography}

%%%%%%%%%%%%%%%%%%%%%%%%%%%%%%%%%%%%%%%%%%%%%%%%%%

%%%%%%%%%%%%%%%%% APPENDICES %%%%%%%%%%%%%%%%%%%%%

\appendix

\section{Magnetic diffusion coefficients for the eight-wave {\sc{flash}} MHD solver}\label{appendix}
The magnetic diffusion coefficients and magnetic pressure gradients obtained for the standard FLASH MHD solver eight-wave are shown in Table \ref{tab:harris_8wave}.

These values were computed with the same procedure and initial conditions described in Section \ref{harris_sect}.
The scaling of $\eta$ for different conditions was also performed in the way described in Section \ref{eta_scale}, obtaining $\log(\eta) \approx 25.49 + 1.04 \log(|\nabla P_{B}|)$.
\begin{table}
 \caption{Magnetic diffusion for the 8-wave MHD solver. The first and second columns represent the simulation name and its resolution respectively. The third and fourth columns
 represent numerical magnetic diffusion coefficients and magnetic pressure gradient magnitudes.}
 \label{tab:harris_8wave}
 \begin{tabular}{cccc}
 \hline
 Simulation & Resolution (\pc) &$\eta \ (\pc^{2}/\Myr)$  & $ |\nabla P_{B}| \ (\mathrm{dyn/cm^{3}})$ \\\hline
 H7 & $\mathrm{7.81\times 10^{-2}}$ &$\mathrm{4.209\times 10^{-3}}$& $\mathrm{3.289 \times 10^{-31}}$\\
 H8 & $\mathrm{3.91\times 10^{-2}}$ &$\mathrm{1.040\times 10^{-3}}$& $\mathrm{3.400 \times 10^{-31}}$\\
 H9 &  $\mathrm{1.95\times 10^{-2}}$ &$\mathrm{2.772\times 10^{-4}}$& $\mathrm{3.389\times 10^{-31}}$\\
 H10 &  $\mathrm{9.75\times 10^{-3}}$ &$\mathrm{7.179\times 10^{-5}}$& $\mathrm{3.378\times 10 ^{-31}}$\\
 H11 & $\mathrm{4.88\times 10^{-3}}$ &$\mathrm{1.819\times 10^{-5}}$& $\mathrm{3.371\times 10^{-31}}$\\\hline
 \end{tabular}
\end{table}

%%%%%%%%%%%%%%%%%%%%%%%%%%%%%%%%%%%%%%%%%%%%%%%%%%

% Don't change these lines
\bsp	% typesetting comment
\label{lastpage}
\end{document}